\documentstyle[twoside,fleqn,espcrc2,psfig]{article}

\newcommand{\AmS}{{\protect\the\textfont2
  A\kern-.1667em\lower.5ex\hbox{M}\kern-.125emS}}

\title{{\it BeppoSAX} detection of the Fe K line in the starburst galaxy NGC253}

\author{S. Mariani$^a$, M. Cappi$^b$, M. Persic$^c$, L. Bassani$^b$, 
        G.G.C. Palumbo$^a$, L. Danese$^d$, A.J. Dean$^e$,
        G. Di Cocco$^b$, A. Franceschini$^f$, L.K. Hunt$^g$, F. Matteucci$^h$, 
        E. Palazzi$^b$, Y. Rephaeli$^i$, P. Salucci$^d$,
        A. Spizzichino$^b$\\
$^{}$\\
$^a$Dpt. of Astronomy, Bologna University, Via Zamboni 33, 40126, Bologna, Italy\\
$^b$ITeSRE/CNR, Via Gobetti 101, 40129, Bologna, Italy\\
$^c$Trieste Astronomical Observatory, Via G.B. Tiepolo 11, 34131, Trieste, Italy\\
$^d$SISSA/ISAS, Via Beirut 4, 34014, Trieste, Italy\\
$^e$Dpt. of Physics, Southampton University, Southampton, SO9 5NH, UK\\
$^f$Dpt. of Astronomy, Padova University, Vicolo dell'Osservatorio 5, 35122, Padova, Italy\\
$^g$CAISMI/CNR, Largo E. Fermi 5, 50125, Firenze, Italy\\
$^h$Dpt. of Astronomy, Trieste University, Via Besenghi, 34131, Trieste, Italy \\
$^i$Tel Aviv University, Tel Aviv, Israel \\}

\begin{document}

\begin{abstract}
Preliminary results obtained from {\it BeppoSAX} observation 
of the starburst galaxy NGC253 are presented.
X-ray emission from the object is clearly extended but most of the emission is 
concentrated on the optical
nucleus. Preliminary analysis of the LECS and MECS data obtained using the central 4' 
region indicates that the continuum is well fitted by two thermal components at 
0.9 keV and 7 keV.
Fe K line at 6.7 keV is detected for the first time in this galaxy; the line
has an equivalent width of $\sim 300$~eV. The line energy and the shape of 
the 2$-$10 keV continuum strongly support thermal origin of the hard X-ray 
emission of NGC253.
From the measurement of the Fe K line the 
abundances can be unambiguously constrained to $\sim~0.25$ the solar value. Other 
lines clearly detected are Si, S and Fe$_{XVIII}$/Ne, in agreement with $ASCA$ results. 
\end{abstract}

\maketitle

\section{Introduction}

Starburst galaxies are spirals (sometimes barred) in which gas is converted to stars at 
rates that could not be sustained over typical galaxy lifetimes. 
Such a phase is thought to represent a significant, if relatively brief,
stage in galactic evolution lasting about 10$^8$~years
(Rieke et al. 1980).

Starbursts tend to be characterized by copious far-infrared (FIR) radiation
from warm interstellar dust heated by massive young stars (Soifer et al. 1986), 
as well as by enhanced radio and X-ray emission.
X-ray emission in starbursts has been attributed to individual point sources,
within the central 10$^3$~pc for nuclear starbursts, 
such as low-mass  X-ray binaries and young supernovae,
and to hot plasma heated by supernova explosions or strong stellar winds from
young massive stars.
Indeed, such hot plasmas have been termed ``superwinds'' (Heckman et al. 1990), 
arising when the supernova rate (e.g. $\sim$0.1 yr$^{-1}$ for NGC253, 
Antonucci \& Ulvestad 1988) and the mass of the gas involved ($\sim$10$^{8}$M$_{\odot}$) 
are high enough to create a shock-heated gas cavity within the galaxy. 
Such cavity could expand, break and then make the hot gas come out 
as superwind. Superwind emission
has been suggested as an explanation for the plume of X-ray, 
discovered by {\it Einstein}, in the northern side of NCG253 (Fabbiano 1988). 

We present here {\it BeppoSAX} observations of a starburst galaxy, NGC253,
that, for the first time, reveal the Fe K line at 6.7 keV.
The detection of this line is fundamental because it constrains the
origin of the X-ray emission, and provides a diagnostic
for plasma temperatures higher than a few keV, and for elemental abundances.
Studies of starburst galaxies are interesting as they help in the understanding  
of the physical processes behind
the high stellar formation rate in the nucleus and in the search for a possible link
between normal galaxies and AGN.  

\section{NGC253}

NGC253 (see table 1 for data) is a nearby edge-on late-typed barred spiral
($i=78^{\circ}.5$, Pence 1981) and represents one
of the archetypical starburst galaxies.
It is one of the brightest infrared sources in the extragalactic sky,
with a 100$\mu$m luminosity of $3.04 \times 10^{10}$~L$_{\odot}$
(Rice et al. 1988), and has been studied extensively at high energies
(Fabbiano \& Trinchieri 1984; Ohashi et al. 1990;  Ptak et al. 1997),
showing a high degree of spectral and spatial complexity at X-ray wavelengths.
\begin{table}[hbt]
\begin{center}
\begin{tabular}{ccccc}
\multicolumn{5}{c}{Table 1: Relevant data of NGC253} \\
\hline
Dec & RA & D$^a$ & d$^b$ & m$_{V}$ \\
\hline
-25d17m18s & 00h47m33s & $\sim$3Mpc & 10'& 8.04 \\
\hline
\end{tabular}
\end{center}
Note: $^a$ see Tully 1988 \\
$^{}$ $^{}$ $^{}$ $^{}$ $^{}$ $^{}$ $^{}$ $^b$ MECS observation  
\end{table}
\section{The {\it BeppoSAX} observation}

NGC253 was observed from November 29 to December 2, 1996 (see table 2). 
In the central 4~arcmin region, we obtained
a LECS count rate of $3.93 \times 10^{-2}$ cts s$^{-1}$ and a MECS 
count rate of $9.23 \times 10^{-2}$ cts s$^{-1}$. Data are characterized,
in the 0.1-10 keV band, by S/N$>$3.  
\begin{table}[hbt]
\begin{center}   
\begin{tabular}{ccc}
\multicolumn{3}{c}{Table 2: Exposure time} \\
\hline
$Instrument$ &  $En. range$ & $Obs. time$ \\
$ $ & $KeV$ & $sec$ \\
\hline
LECS & 0.1-4 & 54689 \\
MECS & 1.3-10 & 113403 \\
\hline
\end{tabular}
\end{center}
\end{table}
The flux observed by {\it BeppoSAX} in the 0.1-2.4 keV energy range,
is $2.36 \times 10^{-12}$~erg~s$^{-1}$~cm$^{-2}$, which is
roughly a factor of two lower than the observed ROSAT flux in the
same energy range (Moran et al. 1996); this lack of agreement may
be attributable to their larger beam, a different background
subtraction technique, or both.
The observed 2-10 keV flux is 
$4.9 \times 10^{-12}$~erg~s$^{-1}$ cm$^{-2}$, consistent with {\it ASCA} 
results (Ptak et al. 1997) and corresponding to
a luminosity of $1.4 \times 10^{40}$~erg~s$^{-1}$.  

\subsection{Spatial and timing analysis}

The source is clearly extended in the {\it BeppoSAX} image in both the 0.1-2 keV and
2-10 keV band. 
Analysis of the resolved emission is postponed to future works;
here we present only the analysis of the unresolved nuclear emission.
In the following N$_{H gal}$=$1.28 \times 10^{20}$ 
cm$^{-2}$ (Dickey \& Lockman 1990) is taken.

No short or long term variability is detected from the present data 
in either energy band: this is consistent with a thermal
origin of the 2-10~keV emission, as discussed below.

\begin{table}[hbt]
\begin{center}
\begin{tabular}{cccc}
&&& \\
\multicolumn{4}{c}{Table 3: Bremsstrahlung + lines model} \\
\hline
$KT_{brem.}$ & $Element$ & $Energy$ & $EW$ \\ 
$keV$ & $ $ & $keV$ & $eV$ \\ 
\hline 
&&& \\
$7.40_{-0.71}^{+0.18}$ & $Fe_{XVIII}/Ne$ & $0.95_{-0.05}^{+0.04}$ 
& $101_{-38}^{+49}$ \\
&&& \\
 & $Si_{XIV}$ & $1.91_{-0.05}^{+0.04}$ & $70_{-29}^{+23}$ \\
&&& \\
 & $S_{XV}$ & $2.42_{-0.06}^{+0.05}$ & $74_{-28}^{+34}$  \\
&&& \\
 & $Fe_{XXV}$ & $6.69_{-0.07}^{+0.07}$ & $310_{-78}^{+119}$ \\
&&& \\
\hline
\end{tabular}
\end{center}
Note: The value of the $A_{LECS}/A_{MECS}$  constant used for this simultaneous fit 
is $0.64_{-0.03}^{+0.04}$
\end{table}

\begin{figure}[htb]
\psfig{file=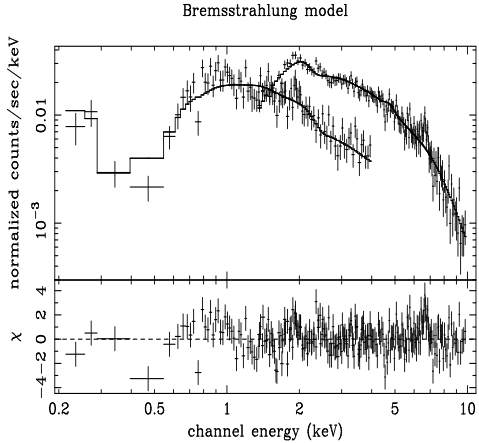,width=7cm,height=6.5cm,angle=-90}
\caption{MECS and LECS fit of a simple Bremsstrahlung model; the lines are 
clearly evident on the continuum}
\label{fig:largenenough}
\end{figure}

\begin{figure}[htb]
\psfig{file=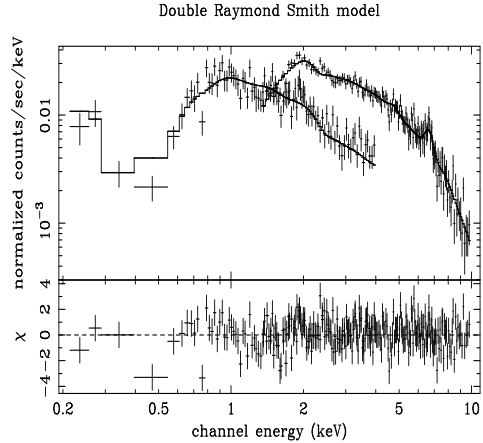,width=7cm,height=6.5cm,angle=-90}
\caption{MECS and LECS fit of a double Raymond Smith model}
\label{fig:toosmall}
\end{figure}

\subsection{Spectral analysis}

At first, we used a bremsstrahlung model plus emission lines to parameterize the 
line energies and
intensities detected with {\it BeppoSAX}. The spectra in both bands were fit 
simultaneously with the relative normalizations free to vary. 
The emission lines are evident in
Fig.1 and the fitted line intensities and gas temperature are given in Table 3.

The {\it BeppoSAX} 2-10 keV continuum clearly requires a thermal model: a hard
power law alone (as allowed by {\it ASCA} data, Ptak et al. 1997) seems to be ruled out 
($\Delta$$\chi^{2}$=49) by the present data.

We find that the spectra are well fitted by a double 
Raymond-Smith model (see Table 4 and Fig.2). The results found using alternative thermal 
models (e.g. Meka and Mekal models in XSPEC) confirm both the temperature 
and abundances found with the Raymond-Smith model. 
The LECS data show a residual excess below 1~keV, requiring a soft
component with KT$<$1~keV, as was also found by Ptak et al. (1997)
with {\it ASCA}.

\begin{table}[hbt]
\begin{center}
\begin{tabular}{cccc}
&&& \\
\multicolumn{4}{c}{Table 4: Double Raymond-Smith model} \\
\hline
$KT_{soft}$ & $Ab_{soft}$ & $KT_{hard}$ & $Ab_{hard}$ \\
$keV$ & $ $ & $keV$ & $ $ \\
\hline 
&&& \\
$0.90_{-0.23}^{+0.19}$ & $\equiv 1$ & $6.52_{-0.50}^{+0.56}$ & 
$0.25_{-0.07}^{+0.08}$ \\
&&& \\
\hline
\end{tabular}
\end{center}
Note: The value of the $A_{LECS}/A_{MECS}$  constant used for this simultaneous fit 
is $0.64_{-0.03}^{+0.04}$
\end{table}

The Fe K line (consistent with emission from Fe$_{XXV}$) at 6.7 keV has been unambiguously
detected for the first time in NGC253.
It is relatively narrow, with
an equivalent width of 310~eV, a value roughly
consistent with the upper limits placed by previous studies 
(Ohashi et al. 1990; Ptak et al. 1997);
we note that a similar emission line was also 
detected in M82 by {\it ASCA} (Ptak et al. 1997).
Other lines clearly detected are Si, S and Fe$_{XVIII}$/Ne (see Table 3), 
in agreement with {\it ASCA} results. The best fit temperature obtained using a 
double Raymond-Smith model is $\sim$6.5 keV, higher than expected but consistent with 
supernovae temperatures. The reliable detection of the Fe K line in NGC253 allows us 
to determine the metallicity of the line-emitting gas, and we find, for the hard 
component, a value of 0.25 solar, again consistent with the sub-solar values, based 
on upper limits, predicted by Ohashi et al. (1990) and Ptak et al. (1997).
However, the quality of the LECS data is too poor to give a reliable estimate
of the metallicity of the soft component.

\section{Conclusions}

The detection of the iron line at 6.7 keV in this starburst galaxy is particulary
interesting: other galaxies, namely LINERs and$/$or low luminosity active galaxies 
tend to show Fe K line 
energies higher than AGN (Iyomoto et al. 1997) raising the question of the nature 
and origin of the line detected in these galaxies in the light of our results. 
The high temperature thermal plasma and the presence of the bump around 
1 keV in the spectra of NGC253, are at present puzzling and require 
further investigation.

\end{document}